\documentclass[letterpaper,onecolumn,11pt]{IEEEtran} 

\usepackage{amsmath,amssymb}
\usepackage{amsthm}
\usepackage{graphics}
\long\def\beginpgfgraphicnamed#1#2\endpgfgraphicnamed{\includegraphics{#1.pdf}} 

\renewcommand{\IEEEQED}{\IEEEQEDopen}

\allowdisplaybreaks

\newcommand{\F}{\mathbb{F}}

\newcommand{\A}{\mathbb{A}}

\newcommand{\ga}{\alpha}
\newcommand{\gb}{\beta}
\newcommand{\gam}{\gamma}

\renewcommand{\phi}{\varphi}

\newcommand{\calL}{\mathcal{L}}
\newcommand{\calO}{\mathcal{O}}

\newcommand{\frakm}{\mathfrak{m}}

\newtheorem{thm}{Theorem}
\newtheorem{lem}[thm]{Lemma}
\newtheorem{prop}[thm]{Proposition}
\newtheorem{cor}[thm]{Corollary}

\newtheorem{ex}{Example}

\newtheorem*{algB}{Algorithm B}

\newtheorem*{algI}{Algorithm I}

\newcommand{\set}[1]{\{#1\}}

\newcommand{\zdeg}{\text{$z$-$\deg$}}

\newcommand{\ideal}[1]{\langle{#1}\rangle}
\newcommand{\ev}{\mathrm{ev}}

\newcommand{\LT}{\mathrm{lt}}
\newcommand{\LC}{\mathrm{lc}}
\newcommand{\ind}{\mathrm{ind}}

\newcommand{\Tr}{\mathrm{Tr}}

\DeclareMathOperator{\mult}{mult}
\DeclareMathOperator{\score}{score}

\begin{document}

\title{Algebraic Soft-Decision Decoding \\ of Hermitian Codes}

\author{Kwankyu~Lee
and Michael E.~O'Sullivan
\thanks{K.~Lee is with the Department of Mathematics, Chosun University, Gwangju 501-759, Korea (e-mail: kwankyu@chosun.ac.kr).}%
\thanks{M.~E.~O'Sullivan is with the Department of Mathematics and Statistics, San Diego State University, San Diego, CA 92182-7720, USA (e-mail: mosulliv@math.sdsu.edu).}%
\thanks{This work was supported by research funds from Chosun University, 2008.}%
}

\maketitle

\begin{abstract}
An algebraic soft-decision decoder for Hermitian codes is presented. We apply Koetter and Vardy's soft-decision decoding framework, now well established for Reed-Solomon codes, to Hermitian codes. First we provide an algebraic foundation for soft-decision decoding. Then we present an interpolation algorithm finding the $Q$-polynomial that plays a key role in the decoding. With some simulation results, we compare performances of the algebraic soft-decision decoders for Hermitian codes and Reed-Solomon codes, favorable to the former.
\end{abstract}

\IEEEpeerreviewmaketitle


\section{Introduction}

Sudan and Guruswami's list decoding of Reed-Solomon codes \cite{sudan1997,guruswami1999} has developed into algebraic soft-decision decoding by Koetter and Vardy \cite{koetter2003}. As Reed-Solomon codes are widely used in coding applications, algebraic soft-decision decoding is regarded as one of the most important developments for Reed-Solomon codes. Hence there have been many subsequent works to make the decoding method efficient and practical \cite{koetter2003b,olshevsky2003,kuijper2004,aleknovich2005,ratnakar2005,ma2007a}. Engineers have proposed fast electronic circuits implementing the algebraic soft-decision decoder \cite{ahmed2004,ma2006,ma2007b}. One may say that the algebraic soft-decision decoding of Reed-Solomon codes is now in a mature state for deployment in applications \cite{gross2006}.

Reed-Solomon codes are the simplest algebraic geometry codes \cite{stichtenoth1993}. Therefore it is natural that the list decoding of Reed-Solomon codes was soon extended to algebraic geometry codes by Shokrollahi and Wasserman \cite{shokrollahi1999} and Guruswami and Sudan \cite{guruswami1999}. However, it seems that no algebraic geometry codes other than Reed-Solomon codes have been considered for algebraic soft-decision decoding. One reason for this unbalanced situation is presumably that the complexity of an algebraic soft-decision decoder for algebraic geometry codes would be prohibitively huge as the complexity for Reed-Solomon codes is already very large. However, algebraic geometry codes have the advantage that they are longer than Reed-Solomon codes over the alphabet of the same size, promising better performance. We may also expect that once we have an explicit formulation of algebraic soft-decision decoding for algebraic geometry codes, some clever ways to reduce the complexity to a practical level may be found, as has happened for Reed-Solomon codes \cite{koetter2003b}.

In this work, we present an algebraic soft-decision decoder for Hermitian codes. Hermitian codes are one of the best studied algebraic geometry codes, and they are often regarded as the first candidate among algebraic geometry codes that could compete with Reed-Solomon codes. To formulate an algebraic soft-decision decoder for Hermitian codes, we basically follow the path set out by Koetter and Vardy for Reed-Solomon codes. Thus there are three main steps of the decoding: the multiplicity assignment step, the interpolation step, and the root-finding step. For the multiplicity assignment step and the root-finding step, we may use algorithms in \cite{koetter2003} and \cite{wu2001}, respectively. Here we focus on the interpolation step, the goal of which is to construct the $Q$-polynomial whose roots give the candidate codewords. As for mathematical contents, this work is an extension of our previous \cite{kwankyu2006c} and \cite{kwankyu2006b}. The core contribution of the present work is an algorithm constructing a set of generators of a certain module from which we extract the $Q$-polynomial using the Gr\"obner conversion algorithm given in \cite{kwankyu2006c}.

In Section 2, we review the definitions of basic concepts and the properties of Hermitian curves and codes. We refer to \cite{fulton1969} and \cite{stichtenoth1993} for the basic theory of algebraic curves and algebraic geometry codes, and \cite{cox1997a} and \cite{atiyah1969} for Gr\"obner bases and commutative algebra. In Section 3, we formulate the algebraic soft-decision decoding of Hermitian codes. We present our interpolation algorithm in Section 4 and a complexity analysis of the decoding algorithm in Section 5. In Section 6, we provide some simulation results of the algebraic soft-decision decoder. As this work is an extension of \cite{kwankyu2006c}, we omitted some proofs that can be found in that work but allowed some similar materials included here for exposition purposes.

\section{Preliminaries}

\subsection{Hermitian curves}

Let $q$ be a prime power, and let $\F$ denote a finite field with $q^2$ elements. The Hermitian curve $H\subset\A^2_\F$ is the affine plane curve defined by the absolutely irreducible polynomial $Y^q+Y-X^{q+1}$ over $\F$. The coordinate ring of $H$ is the integral domain
\[
	R=\F[X,Y]/\ideal{Y^q+Y-X^{q+1}}=\F[x,y],
\]
with $x$ and $y$ denoting the residue classes of $X$ and $Y$, respectively. Note that every element of $R$ can be written uniquely as a polynomial of $x$ and $y$ with $y$-degree less than $q$, as we have $y^q+y-x^{q+1}=0$. So $R$ is also a free module of rank $q$ over $\F[x]$. The function field $K(H)$ is the quotient field of $R$.  

For each  $\ga\in\F$, there are exactly $q$ elements $\gb\in\F$ such that $\Tr_{\F/\F_q}(\gb)=\gb^q+\gb=\ga^{q+1}$. Therefore there are $q^3$ rational points $P_1,P_2,\dots,P_n$ of $H$ with $n=q^3$, which can be grouped into $q^2$ classes of $q$ points with the same $x$-coordinates. A rational point $P$ of $H$ is associated with a maximal ideal $\frakm_P=\set{f\in R\mid f(P)=0}$, and the local ring $\calO_P$ of $H$ at $P$ is the localization of $R$ at $\frakm_P$. For a nonzero $f\in R$, the valuation $v_P(f)$ is the largest integer $r$ such that $f\in\frakm_P^r$.

The projective closure of $H$ is a smooth curve with a unique rational point $P_\infty$ at infinity. The functions $x$ and $y$ on $H$ have poles at $P_\infty$ of orders $q$ and $q+1$, respectively, that is, $v_{P_\infty}(x)=-q$ and $v_{P_\infty}(y)=-q-1$. The genus of $H$ is given by $g=q(q-1)/2$. It is well known that the number of rational points of the curve $H$ attains the maximum value possible for the genus and the size of the base field.

\subsection{Hermitian codes}

For $u\ge 0$, the $\F$-linear space $\calL(uP_\infty)=\set{f\in K(H)\mid (f)+uP_\infty\ge 0}$ has a basis consisting of $x^iy^j$ for $0\le i$, $0\le j\le q-1$, and $qi+(q+1)j\le u$. Therefore
\[
	R=\bigcup_{u=0}^\infty\calL(uP_\infty)=\bigoplus_{\substack{0\le i \\ 0\le j\le q-1}}\!\!\F\cdot x^iy^j.
\]

Recall that the Hamming space $\F^n$ is an $\F$-linear space with the Hamming distance function $d$. For $1\le i\le n$, let $P_i=(\ga_i,\gb_i)$. The evaluation map $\ev:R\to\F^n$ defined by
\[
	\phi\mapsto(\phi(P_1),\phi(P_2),\dots,\phi(P_n))
\]
is a linear map over $\F$. We now fix a positive integer $u$. The Hermitian code $C_u$ is defined to be the image of $\calL(uP_\infty)$ by the evaluation map. If $u<n$, then $\ev$ is injective on $\calL(uP_\infty)$, and the dimension of $C_u$ is equal to $\dim_\F(\calL(uP_\infty))$, which is $u+1-g$ for $u\ge 2g-1$ by the Riemann-Roch theorem. Note also that the minimum distance of $C_u$ is at least $n-u$.

Define
\[
	h_i=-\frac{(x^{q^2}-x)(y^q+y-\gb_i^q-\gb_i)}{(x-\ga_i)(y-\gb_i)}\in R
\]
for $1\le i\le n$. For a vector $v=(v_1,v_2,\dots,v_n)\in\F^n$, define
\[
	h_v=\sum_{i=1}^n v_ih_i.
\]
We can easily prove that $h_i(P_j)=1$ if $j=i$, and $0$ otherwise. Therefore $\ev(h_v)=v$ for all $v\in\F^n$. 

\begin{ex}
Let $q=2$ and $\F_4=\set{0,1,\ga,\ga^2}$. We consider the Hermitian curve $H$ defined by $Y^2+Y+X^3$ over $\F_4$. There are $8$ rational points on $H$, 
\[
	(0,0),(0,1),(1,\ga),(1,\ga^2),(\ga,\ga),(\ga,\ga^2),(\ga^2,\ga),(\ga^2,\ga^2).
\]
Let $u=4$. The linear space $\calL(4P_\infty)$ is spanned by the basis $\set{1,x,y,x^2}$. Hermitian code $C_4$ is a linear code over $\F_4$ of length $8$ and dimension $4$. We use the following generator matrix for encoding
\[
G=\begin{bmatrix}
 1 & 0 & 0 & 1 & 0 & 1 & \ga^2 & \ga \\
 0 & 1 & 0 & 1 & 0 & 1 & \ga & \ga^2 \\
 0 & 0 & 1 & 1 & 0 & 0 & 1 & 1 \\
 0 & 0 & 0 & 0 & 1 & 1 & 1 & 1
\end{bmatrix}.
\]
Note that the positions $1,2,3,5$ form an information set of $G$. Our message is $(1,\ga^2,0,\ga)$, which is encoded into the codeword
\[
	(1,\ga^2,0,\ga)G=(1,\ga^2,0,\ga,\ga,0,0,\ga).
\]
The functions $h_i$ are as follows:
\begin{align*}
	h_1&=(x^3+1)y+x^3+1,\\
	h_2&=(x^3+1)y,\\
	h_3&=(x^3+x^2+x)y+\ga^2x^3+\ga^2x^2+\ga^2x,\\
	h_4&=(x^3+x^2+x)y+\ga x^3+\ga x^2+\ga x,\\
 	h_5&=(x^3+\ga x^2+\ga^2x)y+\ga^2x^3+x^2+\ga x,\\
	h_6&=(x^3+\ga x^2+\ga^2x)y+\ga x^3+\ga^2x^2+x,\\
	h_7&=(x^3+\ga^2x^2+\ga x)y+\ga^2x^3+\ga x^2+x,\\
	h_8&=(x^3+\ga^2x^2+\ga x)y+\ga x^3+x^2+\ga^2x.
\end{align*}
We will continue this example throughout.
\end{ex}

\subsection{Local multiplicity of curves on a surface }

The smooth surface $V=H\times\A_\F^1$ has the coordinate ring $A(V)=R\otimes\F[z]=R[z]$. The function field $K(V)$ is the quotient field of $A(V)$. A rational point $S$ of $V$ is a pair $(P_i,\gamma)$ with $1\le i\le n$ and $\gamma\in\F$, and is associated with a maximal ideal $\frakm_S=\set{f\in A(V)\mid f(S)=0}$. The local ring $\calO_S$ of $V$ at $S$ is the localization of $A(V)$ at $\frakm_S$. A nonzero function $f\in A(V)$ defines a curve on the surface $V$. The multiplicity of $f$ at a rational point $S$, denoted $\mult_S(f)$, is the largest integer $r$ such that $f\in\frakm_S^r$. We note the following properties of multiplicity on the surface $V$. Let $P$ be a rational point of $H$.

\begin{itemize}
\item[(i)] If $f\in R$, then $v_P(f)=\mult_{(P,\gamma)}(f)$ for every $\gamma\in\F$. 
\item[(ii)] For $f\in R$, $\mult_{(P,\gamma)}(z-f)=1$ if $f(P)=\gamma$, and $0$ otherwise. 
\item[(iii)]
For $f,g\in A(V)$, $\mult_S(fg)=\mult_S(f)+\mult_S(g)$ for every rational point $S$ of $V$.
\end{itemize}

\section{Algebraic soft-decision decoding}

Suppose that some codeword of $C_u$ was sent through a noisy channel. The output of the channel is some probabilistic information, for each location  $1\le i \le n$, of the plausibility of each $\gamma\in\F$. The multiplicity assignment step translates the information to a doubly indexed list 
\[
	M=[m_{i\gamma}\mid 1\le i\le n, \gamma\in\F]
\]
of nonnegative integers, where we regard $m_{i\gamma}$ as assigned to the point $(P_i,\gamma)\in H\times\A^1_\F$. The integer value $m_{i\gamma}$ would be chosen roughly proportional to the plausibility of the symbol $\gamma$ according to the channel output. We call $M$ the multiplicity matrix. 

Corresponding to $M$, define 
\[
	I_M=\set{f\in R[z]\mid\text{$\mult_{(P_i,\gamma)}(f)\ge m_{i\gamma}$ for $1\le i\le n, \gamma\in\F$}},
\]
an ideal of $R[z]$. We call $I_M$ the interpolation ideal. Note that by definition
\[
	I_M=\bigcap_{1\le i\le n, \gamma\in\F}\frakm_{(P_i,\gamma)}^{m_{i\gamma}}.
\]

For a vector $v=(v_1,v_2,\dots,v_n)\in\F^n$, the score of $v$ with respect to $M$ is defined as
\[
	\score_M(v)=\sum_{i=1}^n m_{i v_i}
\]
Hence $\score_M(v)$ is also the sum of the multiplicities of the points through which the curve $z-h_v$ passes. The task of the algebraic soft-decision decoder is to find the codeword that has the best score with respect to $M$. This codeword is presumed to be the most likely to have been sent, given the channel output.

\begin{ex}[continued]\label{djcqoe}
Suppose that the codeword in the previous example is sent through a noisy channel and received data gives rise to the matrix of the plausibilities of symbols 
\[
\begin{bmatrix}
0.604 & 0.001 & 0.171 & 0.001 & 0.567 & 0.949 & 0.997 & 0.486 \\
0.396 & 0.158 & 0.760 & 0.000 & 0.103 & 0.010 & 0.003 & 0.022 \\
0.000 & 0.005 & 0.013 & 0.985 & 0.279 & 0.041 & 0.000 & 0.470 \\
0.000 & 0.836 & 0.056 & 0.014 & 0.051 & 0.000 & 0.000 & 0.022
\end{bmatrix}
\]
which is then translated to the multiplicity matrix
\[
M=\begin{bmatrix}
3 & 0 & 0 & 0 & 2 & 4 & 5 & 2 \\
2 & 0 & 3 & 0 & 0 & 0 & 0 & 0 \\
0 & 0 & 0 & 5 & 1 & 0 & 0 & 2 \\
0 & 4 & 0 & 0 & 0 & 0 & 0 & 0
\end{bmatrix},
\]
where the rows are indexed by $\gamma=0,1,\ga,\ga^2$ from top to bottom. Note that neither of the vectors $(0,\ga^2,1,\ga,0,0,0,0)$ and $(0,\ga^2,1,\ga,0,0,0,\ga)$ that have the best score with respect to $M$ is a codeword of $C_4$.
\end{ex}

\begin{lem}\label{lemabcd}
Let $M=[m_{i\gamma}]$ be the multiplicity matrix. Then 
\[
	\dim_{\F} R[z]/I_M=\sum_{i=1}^n\sum_{\gamma\in\F}\binom{m_{i\gamma}+1}{2}.
\]
\end{lem}

\begin{IEEEproof}
Since $I_M$ is a zero-dimensional ideal,
\[
\begin{split}
	\dim_{\F} R[z]/I_M&=\sum_{i=1}^n\sum_{\gamma\in\F}\dim_{\F}\calO_{(P_i,\gamma)}/I_M\calO_{(P_i,\gamma)}\\
		&=\sum_{i=1}^n\sum_{\gamma\in\F}\dim_{\F}\hat\calO_{(P_i,\gamma)}/I_M\hat\calO_{(P_i,\gamma)},
\end{split}
\]
where $\hat\calO$ denotes the completion of the local ring. If $t$ is a uniformizing parameter of $P_i$ and $s=z-\gamma$, then $\hat\calO_{(P_i,\gamma)}$ is isomorphic to $\F[[s,t]]$. So $\hat\calO_{(P_i,\gamma)}/I_M\hat\calO_{(P_i,\gamma)}$ is isomorphic to $\F[[s,t]]/(s,t)^{m_{i\gamma}}$. The conclusion follows.
\end{IEEEproof}

\begin{lem}
Let $\mu\in R$ with $v=\ev(\mu)$. Then
\[
	\dim_\F R[z]/(I_M+\ideal{z-\mu})=\score(v).
\]
\end{lem}

\begin{IEEEproof}
As in the previous proof,
\[
	\dim_{\F} R[z]/(I_M+\ideal{z-\mu})
	=\sum_{i=1}^n\sum_{\gamma\in\F}\dim_{\F}\hat\calO_{(P_i,\gamma)}/(I_M+\ideal{z-\mu})\hat\calO_{(P_i,\gamma)}.
\]
Let $t$ be a uniformizing parameter of $P_i$ and $s=z-\gamma$ again. We find that if $\mu(P_i)=\gamma$, then $\hat\calO_{(P_i,\gamma)}/(I_M+\ideal{z-\mu})\hat\calO_{(P_i,\gamma)}$ is isomorphic to $\F[[s,t]]/(\ideal{s,t}^{m_{i\gamma}}+\ideal{s-tu})=\F[[t]]/\ideal{t^{m_{i\gamma}}}$, but collapses to the zero ring otherwise. Here $u$ is some unit in $\hat\calO_{(P_i,\gamma)}$. The conclusion follows.
\end{IEEEproof}

For $f=\psi_az^a+\dots+\psi_1z+\psi_0\in R[z]$ with $\psi_i\in R$, the $u$-weighted degree of $f$ is defined as
\[
	\deg_u(f)=\max_{0\le i\le a}(-v_{P_\infty}(\psi_i)+ui).
\]
For $f\in R[z]$ and $\phi\in R$, denote by $f(\phi)$ the element in $R$ that is obtained by substituting $z$ with $\phi$ in $f$. Observe that if $\phi\in\calL(uP_\infty)$, then $-v_{P_\infty}(f(\phi))\le\deg_u(f)$. The algebraic soft-decision decoding of Hermitian codes rests upon the following

\begin{prop}\label{thmakwe}
Suppose $f\in I_M$ is nonzero. If a codeword $c=\ev(\mu)$ of $C_u$ with $\mu\in\calL(uP_\infty)$ satisfies
\[
	\score_M(c)>\deg_u(f),
\]
then $f(\mu)=0$.
\end{prop}

\begin{IEEEproof}
Assume that $f(\mu)$ is not zero in $R$. Then
\[
\begin{split}
	\deg_u(f)	&\ge-v_{P_\infty}(f(\mu))\\
				&=\dim_\F(R/f(\mu)) \\
				&=\dim_\F(R[z]/\ideal{f,z-\mu})\\
				&\ge\dim_\F(R[z]/(I_M+\ideal{z-\mu})=\score(c).
\end{split}
\]
For the first equality, see Lemma 5 in \cite{kwankyu2006c}. This implies that if $\score_M(c)>\deg_u(f)$, we must have $f(\mu)=0$. 
\end{IEEEproof}

In the interpolation step, the decoder picks a polynomial $f\in I_M$. Then by Proposition \ref{thmakwe}, all codewords whose score with respect to $M$ is big enough can be obtained from the roots of $f$ over $R$. Thus the decoder can find among the candidates the codeword that has the best score with respect to $M$. It should be noted that for the best performance of algebraic soft-decision decoding, it is crucial for the decoder to find a polynomial in $I_M$ with the smallest $u$-weighted degree. Having the same weighted degree, the one with smaller degree in $z$ is preferred because this reduces the work of the root-finding step. Here the idea of Gr\"obner bases is relevant. 

We call the elements in the set 
\[
	\Omega=\set{x^iy^jz^k\mid 0\le i, 0\le j\le q-1, 0\le k}
\]
monomials of $R[z]$. Recall that every element of $R[z]$ can be written as a unique linear combination over $\F$ of monomials of $R[z]$. Note that 
\[
	\deg_u(x^iy^jz^k)=qi+(q+1)j+uk.
\]
For two monomials $x^{i_1}y^{j_1}z^{k_1}$, $x^{i_2}y^{j_2}z^{k_2}$ in $\Omega$, we declare 
\[
	x^{i_1}y^{j_1}z^{k_1}>_u x^{i_2}y^{j_2}z^{k_2}
\] 
if $\deg_u(x^{i_1}y^{j_1}z^{k_1})>\deg_u(x^{i_2}y^{j_2}z^{k_2})$ or $k_1>k_2$ when tied. It is easy to verify that $>_u$ is a total order on $\Omega$. Notions such as the leading term and the leading coefficient of $f\in R[z]$ are defined in the usual way. For $f\in R[z]$, the $z$-degree of $f$, written $\zdeg(f)$, is the degree of $f$ as a polynomial in $z$ over $R$.

Now we define the $Q$-polynomial of $I_M$ as the unique, up to a constant multiple, element in $I_M$ with the smallest leading term with respect to $>_u$. By the definition, the $Q$-polynomial is an element of $I_M$ with the smallest $u$-weighted degree, and moreover it has the smallest $z$-degree among such elements. Therefore we may say that the $Q$-polynomial is an optimal choice for the interpolation step.

The last step of algebraic soft-decision decoding is to compute roots of the $Q$-polynomial over $R$ or the function field $K(H)$. Only those roots that belong to $\calL(uP_\infty)$ yield candidate codewords. If the list of the candidate codewords is empty, the decoder may declare decoding failure or resort to hard-decision decoding directly from the channel output. If there are several codewords in the list, then the decoder chooses the codeword that has the best score, and outputs the received message by projecting the codeword on the information set.

\begin{ex}[continued]\label{cjops}
The $Q$-polynomial of $I_M$ 
\begin{equation}\label{dwkxzs}
\begin{split}
Q&=z^5+(\ga^2x^3+\ga xy+x^2+\ga y)z^4+(\ga x^4y+\ga^2x^5+x^3+xy+x^2+y+\ga^2x+1)z^3\\
&\quad+(\ga^2x^6y+\ga x^7+\ga^2x^5y+\ga^2x^6+\ga^2x^4y+\ga^2x^5+\ga^2x^3y+x^4+x^3+\ga xy+\ga^2x)z^2\\
&\quad+(x^8y+\ga^2x^9+\ga x^8+\ga^2x^7+x^6+\ga x^4y+x^5+\ga x^3y+\ga x^4+\ga^2x^3+\ga^2xy+\ga x^2+\ga^2y)z\\
&\quad+\ga^2x^{11}+x^{10}+x^8y+\ga x^9+\cdots+\ga^2x^2y+x^3+\ga^2xy+y
\end{split}
\end{equation}
is obtained by the interpolation algorithm in the next section. It turns out the $Q$-polynomial has the factorization
\[
\begin{split}
Q&=\Bigl(z+x^2+\ga^2y+x\Bigr)\Bigl(z+\ga^2x^2+\ga y+x+1\Bigr)\Bigl(z^3+(\ga^2x^3+\ga xy+\ga^2x^2+\ga^2y+1)z^2\\
&\quad +(\ga x^4y+\ga x^5+\ga x^2y+\ga x^3+\ga y+\ga x)z+x^7+\ga^2x^6+x^3y+x^4+\ga x^2y+\ga^2x^3+\ga xy+\ga^2y\Bigr).
\end{split}
\]
Therefore a root-finding algorithm will output two roots. The first root $x^2+\ga^2y+x$ gives the codeword
\[
	c_1=(0,\ga^2,1,\ga,0,\ga^2,0,\ga^2)
\]
whose score is $22$ while the second root $\ga^2x^2+\ga y+x+1$ gives the codeword
\[
	c_2=(1,\ga^2,0,\ga,\ga,0,0,\ga)
\]
whose score is $23$. Therefore the decoder chooses $c_2$, and the received message is
\[
	(1,\ga^2,0,\ga),
\]
which is the correct sent message.
\end{ex}

We will need upper bounds on the $u$-weighted degree and the $z$-degree of the $Q$-polynomial of $I_M$. Let $Q$ denote the $Q$-polynomial of $I_M$. 

\begin{prop}\label{propdkwd}
If $A\subset\Omega$ is a finite set of monomials of $R[z]$ such that 
\[
	|A|>\sum_{i=1}^n\sum_{\gamma\in\F}\binom{m_{i\gamma}+1}{2},
\]
then there is a set of coefficients $c_\phi\in\F$ such that $0\neq \sum_{\phi\in A}c_\phi\phi\in I_M$.
\end{prop}

\begin{IEEEproof}
Lemma \ref{lemabcd} implies that monomials in $A$ are linearly dependent over $\F$ in $R[z]/I_M$. On the other hand, they are linearly independent over $\F$ in $R[z]$. 
\end{IEEEproof}

In a table, we arrange monomials of $R[z]$ such that the monomials in the same column have the same $u$-weighted degree and the monomials in the same row have the same $z$-degree. Let weighted degrees increase from left to right and $z$-degrees from bottom to top.

\begin{ex}[continued]
Note that $\deg_u(x^iy^jz^k)=2i+3j+4k$. So we have the following table
\[
\begin{array}{c|*{4}{c}|*{4}{c}|*{4}{c}|*{4}{c}|*{3}{c}}
	3& &    & & &&&   &  &   &        &    &    &z^3 &\bigcirc&\cdots\\
	2& &    & & &&&   &  &z^2&\bigcirc&xz^2&yz^2&x^2z^2&xyz&\cdots\\
	1& &    & & &z&\bigcirc&xz &yz&x^2z&xyz&x^3z&x^2yz&x^4z&x^3yz&\cdots\\
	0&1&\bigcirc&x&y&x^2&xy&x^3&x^2y&x^4&x^3y&x^5&x^4y&x^6&x^5y&\cdots\\\hline
	&0&1&2&3&4&5&6&7&8&9&10&11&12&13
\end{array}
\]
The symbol $\bigcirc$ indicates that there is no monomial for the position.
\end{ex}

The table of monomials of $R[z]$ suggests the following formula. Let $G(i)=0$ if $i$ is a Weierstrass gap at $P_\infty$, and $1$ otherwise. Note that $G(i)=1$ for $i\ge 2g$. The number of monomials with $u$-weighted degree $i$ is 
\[
	C(i)=\sum_{j=0}^{\lfloor i/u\rfloor}G(i-uj).
\]
Let $w$ be the smallest integer such that
\[
	N=1+\sum_{i=1}^n\sum_{\gamma\in\F}\binom{m_{i\gamma}+1}{2}\le\sum_{i=0}^wC(i).
\]
Let $l=\lfloor w/u\rfloor$. Then the $u$-weighted degrees and the $z$-degrees of monomials up to the $N$th monomial are not greater than $w$ and $l$, respectively. Now Proposition \ref{propdkwd} implies $\deg_u(Q)\le w$ and $\zdeg(Q)\le l$. 

\begin{ex}[continued]
$G(0)=1,G(1)=0$, and $G(i)=1$ for $i\ge 2$ since $g=1$. So we have
\[
	\begin{array}{c|*{12}{c|}c}
		i   & 0 & 1 & 2 & 3 &4 &5 &\cdots &21 &22 &23 &24 &25 &\cdots\\\hline
	  C(i)& 1 & 0 & 1 & 1 &2 &1 &\cdots &5  &6  &6  &7  &6  &\cdots\\\hline
	\sum_{j=0}^i C(j)	& 1 & 1 & 2 & 3 &5 &6 &\cdots &66&72&78&85&91&\cdots
	\end{array}
\]
For our $M$, $N=76$. Therefore $w=23$, $l=5$. Hence $\deg_u(Q)\le 23$ and $\zdeg(Q)\le 5$. 
\end{ex}

\section{An interpolation algorithm}

Let $l$ be a positive integer such that $\zdeg(Q)\le l$. Define
\[
	R[z]_l=\set{f\in R[z]\mid \zdeg(f)\le l}.
\]
Note that $R[z]_l$ is a free module over $R$ of rank $l+1$ with a free basis $1,z,\dots,z^l$. Define 
\[
	I_{M,l}=I_M\cap R[z]_l.
\]
Clearly $I_{M,l}$ is a submodule of $R[z]_l$ over $R$.

Recall that the ring $R=\F[x,y]$ is in turn a free module over $\F[x]$ of rank $q$, with a free basis $\set{1,y,\dots,y^{q-1}}$. So we may view $R[z]_l$ as a free module of rank $q(l+1)$ over $\F[x]$ with a free basis $\set{y^jz^i\mid 0\le i\le l, 0\le j\le q-1}$. The elements of $\Omega\cap R[z]_l$ will be called monomials of $R[z]_l$. It is clear that the total order $>_u$ is precisely a monomial order on the free module $R[z]_l$ over $\F[x]$.
We also view $I_{M,l}$ as a submodule of the free module $R[z]_l$ over $\F[x]$. It is immediate that the $Q$-polynomial of $I_M$ is also the element of $I_{M,l}$ with the smallest leading term with respect to $>_u$. As a consequence of the definition of Gr\"obner bases, $Q$ occurs as the smallest element in any Gr\"obner basis of the module $I_{M,l}$ over $\F[x]$ with respect to $>_u$. 

\subsection{Generators of the module $I_{M,l}$ over $R$}\label{subsecA}

We begin with 

\begin{prop}
Let $M=[m_{i\gamma}]$ be a doubly indexed list of nonnegative integers. For each $i=1,2,\dots, n$, let $n_i=\max_{\gamma\in\F}m_{i\gamma}$, and let $N=[n_i]$. For each $i$ with $n_i>0$, let $\gamma_i$ be such that $m_{i\gamma_i}=n_i$. Let $M'=[m_{i\gamma}']$ where $m_{i\gamma}'=m_{i\gamma}$ for $\gamma\neq\gamma_i$ and $m_{i\gamma_i}'=m_{i\gamma_i}-1$. Then as a module over $R$,
\[
	I_M=(z-h)I_{M'}+J_N
\]
where $J_N=\set{f\in R\mid v_{P_i}(f)\ge n_i}$ and $h\in R$ is such that $h(P_i)=\gamma_i$.
\end{prop}

\begin{IEEEproof}
By the properties (i), (ii), (iii) of local multiplicity, it is clear that $(z-h)I_{M'}+J_N\subset I_M$. To show the reverse inclusion, let $f\in I_M$. We can write $f=(z-h)g+r$ for some $g\in R[z]$ and $r\in R$. Let $S=(P_i,\gamma_i)$. If $t$ is a uniformizing parameter of $P_i$, then $t$ and $z-\gamma_i$ form a system of parameters of $\calO_S$. Recall that the completion $\hat\calO_S$ is isomorphic to the power series ring $\F[[z-\gamma_i,t]]$. Now if $v_{P_i}(r)=c$, then in $\hat\calO_S$,
\[
	f=(z-\gamma_i)g+t^cu
\]
for some unit $u$ in $\hat\calO_S$. Since $f\in\ideal{z-\gamma_i,t}^{n_i}$ and $z-\gamma_i,t$ are algebraically independent over $\F$, we see that $c\ge n_i$. Then as this is true for all $1\le i\le n$, it follows that $r\in J_N$. Hence $f-r=(z-h)g\in I_M$. Again by the properties of local multiplicity, $g\in I_{M'}$. Thus we showed the reverse inclusion.
\end{IEEEproof}

Recall the multiplicity matrix $M=[m_{i\gamma}]$. Let $l_\mathrm{max}=\max_i\set{\sum_{\gamma\in\F}m_{i\gamma}}$. Initially let $m_{i\gamma}^{(0)}=m_{i\gamma}$ and $n_i^{(0)}=\max_{\gamma\in \F} \set{m_{i\gamma}}$. Proceed inductively for $s=0,1,\dots,l_\mathrm{max}-1$. Choose $\gamma_i$ such that $m_{i\gamma_i}^{(s)}=n_i^{(s)}$ if $n_i^{(s)}>0$. Let $h^{(s)}\in R$ such that $h^{(s)}(P_i)=\gamma_i$. Let
\begin{align*}
	m_{i\gam}^{(s+1)} &= 	
		\begin{cases}
			m_{i\gam}^{(s)}-1&\text{if $\gam=\gam_{i}$}, \\
			m_{i\gam}^{(s)} &\text{if $\gam\not=\gam_{i}$},
	\end{cases}\\
	n_i^{(s+1)}&= \max_{\gam\in \F}m_{i\gam}^{(s+1)}.
\end{align*}
Now let $M^{(s)}=[m_{i\gam}^{(s)}]$ and $N^{(s)}=[n_i^{(s)}]$. Observe $m_{i\gamma}^{(l_\mathrm{max})}=0$ for all $i,\gamma$, and therefore $I_{M^{(l_\mathrm{max})}}=R[z]$. By induction, we get

\begin{cor}
For $0\le l$, 
\[ 
	I_{M,l} = \sum_{s=0}^lJ_{N^{(s)}} \prod_{0\leq r<s} (z-h^{(r)})
\]
as a module over $R$. Here $J_{N^{(s)}}=R$ and $h^{(s)}=0$ for $s\ge l_\mathrm{max}$. 
\end{cor}

\subsection{Computing generators of the module $I_{M,l}$ over $\F[x]$}

We may view the ideal $J_N=\set{f\in R\mid v_{P_i}(f)\ge n_i}$ as a module over $\F[x]$. Indeed $J_N$ is a free module of rank $q$ over $\F[x]$. Thus we obtain 

\begin{algB}
The input is an $n\times q^2$ matrix $M=[m_{i\gamma}]$ of nonnegative integers. The output is the generators  $\set{g_{s,t}\mid 0\le s\le l, 0\le t\le q-1}$ of $I_{M,l}$ as a module over $\F[x]$. Repeat steps B1 and B2 for $s=0,1,\dots,l$.
\begin{itemize}
\item[B1.] Let $n_i=\max_{\gamma \in \F}m_{i\gamma}$ for $1\le i\le n$. Let $L=\set{1\le i\le n\mid n_i\ge 1}$. For each $i\in L$, let $\gamma_i\in\F$ be such that $n_i=m_{i\gamma_i}$.
Set
\begin{equation}\label{eqjcjsq}
	g_{s,t}\leftarrow\eta_t\prod_{0\le r<s}(z-h^{(r)}), 
\end{equation}
for $0\le t\le q-1$, where $\set{\eta_0,\eta_1,\dots,\eta_{q-1}}$ is a set of generators of $J_{N^{(s)}}=\set{f\in R\mid v_{P_i}(f)\ge n_i}$ as a module over $\F[x]$. When $L$ is empty, $J_N=R$ so $\eta_t=y^t$.
\item[B2.] Set
\[
	h^{(s)} \leftarrow\sum_{i\in L}\gamma_ih_i
\]
and for $1\le i\le n,\gamma\in\F$, set
\[
	m_{i\gamma}\leftarrow
		\begin{cases}
			m_{i\gamma}-1&\text{if $\gamma=\gamma_i$,}\\
			m_{i\gamma}&\text{otherwise.}
		\end{cases}
\]
\end{itemize}
\end{algB}

Notice that if we compute $\eta_t$ by the method in the following subsection, $g_{s,t}$ has leading term $y^tz^s$ with respect to lex order $x<y<z$.

\begin{ex}[continued]
We continue from Example \ref{djcqoe}. We show the first few steps to compute a set of generators of $I_{M,l}$ with $l=5$ using Algorithm B. Initially $s=0$. Then 
\[
	n_1=3, n_2=4, n_3=3, n_4=5, n_5=2, n_6=4, n_7=5, n_8=2.
\]
As we compute in Example \ref{jwqdd},
\[
	J_{N^{(0)}}=\langle x^{18}+\ga x^{17}+\ga^2x^{16}+x^6+\ga x^5+\ga^2x^4,
	(x^{10}+x^9+x^4+x^3)y+\ga^2x^{17}+x^{16}+\cdots+x^3\rangle
\]
as a module over $\F[x]$. So we set
\begin{align*}
	g_{0,0}&=x^{18}+\ga x^{17}+\ga^2x^{16}+x^6+\ga x^5+\ga^2x^4,\\
	g_{0,1}&=(x^{10}+x^9+x^4+x^3)y+\ga^2x^{17}+x^{16}+\cdots+x^3.
\end{align*}
In step B2, we compute (setting $\gamma_8=0$ arbitrarily)
\[
\begin{split}
	h^{(0)}&=0h_1+\ga^2 h_2+1 h_3+\ga h_4+0 h_5+0 h_6+0 h_7+0 h_8\\
	&=\ga^2x^2y+\ga^2xy+\ga^2y.
\end{split}
\]
Now the matrix of $m_{i\gamma}$ is
\[
\begin{bmatrix}
2 & 0 & 0 & 0 & 1 & 3 & 4 & 1 \\
2 & 0 & 2 & 0 & 0 & 0 & 0 & 0 \\
0 & 0 & 0 & 4 & 1 & 0 & 0 & 2 \\
0 & 3 & 0 & 0 & 0 & 0 & 0 & 0
\end{bmatrix}.
\]

Going on to $s=1$, we have 
\[
	n_1=2, n_2=3, n_3=2, n_4=4, n_5=1, n_6=3, n_7=4, n_8=2.
\]
Then
\[
	J_{N^{(1)}}=\ideal{x^{14}+\ga x^{13}+\ga^2x^{12}+\cdots+\ga x^4+\ga^2x^3,\
	\ga x^{13}+\ga^2x^{12}+\ga^2x^{11}+\cdots+\ga x^3+\ga^2x^2}
\]
as a module over $\F[x]$. Hence
\begin{align*}
	g_{1,0}&=(x^{14}+\ga x^{13}+\ga^2x^{12}+\cdots+\ga x^4+\ga^2x^3)z
	+(\ga^2x^{16}+\ga x^{15}+\ga^2x^4+\ga x^3)y\\
	g_{1,1}&=(x^7+\ga x^6+\ga^2x^5+x^4+\ga x^3+\ga^2x^2)yz\\
		&+(\ga x^{13}+\ga^2x^{12}+\ga^2x^{11}+\cdots+\ga x^3+\ga^2x^2)z \\
		&+(x^{15}+\ga^2x^{14}+x^{13}+x^{10}+x^9+\ga^2x^8+x^7+x^4)y\\
		&+\ga^2x^{12}+\ga x^{11}+\ga^2x^6+\ga x^5
\end{align*}
Now $h^{(1)}=\ga x^3y+\ga x^2y+\ga^2x^3+\ga x^2+\ga^2y+x$ and the matrix of $m_{i\gamma}$ is
\[
\begin{bmatrix}
1 & 0 & 0 & 0 & 0 & 2 & 3 & 1 \\
2 & 0 & 1 & 0 & 0 & 0 & 0 & 0 \\
0 & 0 & 0 & 3 & 1 & 0 & 0 & 1 \\
0 & 2 & 0 & 0 & 0 & 0 & 0 & 0
\end{bmatrix}.
\]
Proceeding this way until $s=5$, we obtain a set of generators of the module $I_{M,l}$. We arrange the coefficients (polynomials in $x$) of the generators in the following matrix
\begin{equation}\label{dwhxas}
\begin{bmatrix}
x^{18}+\cdots & 0 & 0 & 0 & \cdots & 0 & 0 & 0 & 0 \\
\ga^2x^{17}+\cdots & x^{10}+\cdots & 0 & 0 & \cdots & 0 & 0 & 0 & 0 \\
0 & \ga^2x^{16}+\cdots & x^{14}+\cdots & 0 & \cdots & 0 & 0 & 0 & 0 \\
\ga^2x^{12}+\cdots & x^{15}+\cdots & \ga x^{13}+\cdots & x^7+\cdots & \cdots & 0 & 0 & 0 & 0 \\
x^{18}+\cdots & \ga^2x^{15}+\cdots & \ga^2x^{13}+\cdots & \ga x^{13}+\cdots & \cdots & 0 & 0 & 0 & 0 \\
\ga x^{17}+\cdots & x^{14}+\cdots & x^{12}+\cdots & \ga^2x^{12}+\cdots & \cdots & 0 & 0 & 0 & 0 \\
\ga x^{17}+\cdots & \ga^2x^{17}+\cdots & x^{15}+\cdots & \ga^2x^{12}+\cdots & \cdots & 0 & 0 & 0 & 0 \\
\ga x^{16}+\cdots & \ga x^{16}+\cdots & \ga^2x^{14}+\cdots & \ga^2x^{11}+\cdots & \cdots & 0 & 0 & 0 & 0 \\
\ga x^{20}+\cdots & \ga^2x^{17}+\cdots & \ga x^{15}+\cdots & \ga^2x^{15}+\cdots & \cdots & x^3+\cdots & 0 & 0 & 0 \\
x^{18}+\cdots & \ga x^{17}+\cdots & \ga^2x^{15}+\cdots & \ga x^{13}+\cdots & \cdots & \ga^2x & 1 & 0 & 0 \\
x^{20}+\cdots & x^{20}+\cdots & \ga x^{18}+\cdots & \ga x^{15}+\cdots & \cdots & x^3+\cdots & x^3+\ga x & 1 & 0 \\
x^{23}+\cdots & x^{19}+\cdots & \ga x^{18}+\cdots & \ga x^{18}+\cdots & \cdots & x^6+\ga x^4 & \ga x^2+x & 0 & 1
\end{bmatrix}
\end{equation}
where the rows are $g_{0,0}$, $g_{0,1}$, $g_{1,0}$, $g_{1,1}$, \dots, $g_{5,0}$, $g_{5,1}$ in this order, and the columns are coefficients of $1$, $y$, $z$, $yz$, $z^2$, $yz^2$, \dots, $z^5$, $yz^5$ in this order. 
\end{ex}

\subsection{Computing generators of $J_N$}\label{subsecB}

We now tackle the task of computing a set of generators of $J_N$ as a module over $\F[x]$. For this, we switch to a different indexing of the rational points of $H$ by grouping the $q^3$ rational points into $q^2$ classes with the same $x$-coordinates. Thus the rational points are $P_{a,b}=(\ga_a,\gb_{a,b})$ for $1\le a\le q^2$ and $1\le b\le q$. Let $\mu_{a,b}=n_i$ if $P_{a,b}$ is the point $P_i$. Also assume that for each $1\le a\le q^2$, we have arranged the index $b$ such that $\mu_{a,b}$ are put in decreasing order, 
\[
	\mu_{a,1}\ge\mu_{a,2}\ge\cdots\ge\mu_{a,q}.
\]
With the new notations, 
\[
	J_N=\set{f\in R\mid\text{$v_{P_{a,b}}(f)\ge\mu_{a,b}$ for $1\le a\le q^2$ and $1\le b\le q$}}.
\]

\begin{prop}\label{prop6}
For $1\le b<c\le q$, suppose that $f_{b,c}\in\F[x]$ satisfy 
\[
	v_{P_{a,b}}(y-f_{b,c})\ge\mu_{a,b}-\mu_{a,c}
\]
for all $1\le a\le q^2$. Define for $c=1,2,\dots,q$, 
\begin{equation}\label{equxcjms}
	g_c=\prod_{1\le a\le q^2}(x-\ga_a)^{\mu_{a,c}}\prod_{1\le b<c}(y-f_{b,c}).
\end{equation}
Then $J_N=\ideal{g_1,g_2,\dots,g_q}$ as a module over $\F[x]$.
\end{prop}

\begin{IEEEproof}
Let $1\le c\le q$. Then for $1\le a\le q^2$ and $1\le b\le q$, 
\[
\begin{split}
	v_{P_{a,b}}(g_c)&=\mu_{a,c}v_{P_{a,b}}(x-\ga_a)+\sum_{1\le b'<c}v_{P_{a,b}}(y-f_{b',c})\\
		&\ge\mu_{a,c}+v_{P_{a,b}}(y-f_{b,c})\ge\mu_{a,b}.
\end{split}
\]
Therefore $g_c\in J_N$. Recall that we may view $R$ as a free module of rank $q$ over $\F[x]$. Let $J$ be the submodule of $R$ generated by $g_1,g_2,\dots,g_q$ over $\F[x]$. Then $R/J$ is isomorphic to 
\[
	\bigoplus_{1\le c\le q}\F[x]/\ideal{\prod_{1\le a\le q^2}(x-\ga_a)^{\mu_{a,c}}}.
\]
Therefore
\[
	\dim_\F R/J=\sum_{1\le c\le q}\dim_\F\F[x]/\ideal{\prod_{1\le a\le q^2}(x-\ga_a)^{\mu_{a,c}}}
		=\sum_{1\le c\le q}\sum_{1\le a\le q^2}\mu_{a,c}.
\]
On the other hand, as $J_N=\bigcap_{P_{a,b}}\frakm_{P_{a,b}}^{\mu_{a,b}}$ by its definition, we have
\[
	\dim_\F R/J_N=\sum_{P_{a,b}}\dim_\F\calO_{P_{a,b}}/\frakm_{P_{a,b}}^{\mu_{a,b}}
		=\sum_{1\le a\le q^2,1\le b\le q}\mu_{a,b}.
\]
Hence $\dim_\F R/J=\dim_\F R/J_N$. Together with $J\subset J_N$, this implies that $J=J_N$.
\end{IEEEproof}

\begin{ex}[continued]\label{jwqdd}
We compute generators $g_1$, $g_2$ of $J_{N^{(0)}}$.  We arrange the points as
\begin{align*}
	P_{1,1}=P_2,&\quad P_{1,2}=P_1,\\
	P_{2,1}=P_4,&\quad P_{2,2}=P_3,\\
	P_{3,1}=P_6,&\quad P_{3,2}=P_5,\\
	P_{4,1}=P_7,&\quad P_{4,2}=P_8,
\end{align*}
so that $\mu_{a,b}$ are in decreasing order,
\begin{align*}
	\mu_{1,1}=4&,\mu_{1,2}=3,\\
	\mu_{2,1}=5&,\mu_{2,2}=3,\\
	\mu_{3,1}=4&,\mu_{3,2}=2,\\
	\mu_{4,1}=5&,\mu_{4,2}=2.
\end{align*}
We will see in the next subsection that
\[
	f_{1,2}=\ga^2x^7+\ga x^6+\ga x^4+x^3+\ga^2x^2+y+\ga^2x+1
\]
satisfies
\begin{align*}
	v_{P_{1,1}}(y-f_{1,2})&\ge 1,\\
	v_{P_{2,1}}(y-f_{1,2})&\ge 2,\\
	v_{P_{3,1}}(y-f_{1,2})&\ge 2,\\
	v_{P_{4,1}}(y-f_{1,2})&\ge 3.
\end{align*}
Therefore 
\begin{align*}
	g_1&=(x-0)^4(x-1)^5(x-\ga)^4(x-\ga^2)^5=x^{18}+\ga x^{17}+\ga^2x^{16}+x^6+\ga x^5+\ga^2x^4,\\
	g_2&=(x-0)^3(x-1)^3(x-\ga)^2(x-\ga^2)^2(y-f_{1,2})\\
	&=(x^{10}+x^9+x^4+x^3)y+\ga^2x^{17}+x^{16}+\ga x^{15}+\ga x^{14}+\ga^2x^{13}+\ga x^{12}+\ga^2x^{11}\\
	&\quad +\ga^2x^{10}+\ga^2x^9+\ga x^8+\ga^2x^7+\ga x^6+\ga x^4+x^3
\end{align*}
generates $J_{N^{(0)}}$ as a module over $\F[x]$.
\end{ex}

\subsection{Computing $y-f_{b,c}$}\label{subsecC}

As $y=x^{q+1}-y^q$, we see that $y=\sum_{i=0}^\infty(-1)^ix^{(q+1)q^i}$ in the completion of the local ring at $(0,0)$. On the other hand, if $(\ga,\gb)$ is a rational point of $H$, then $x\mapsto x-\ga$, $y\mapsto y-\ga^q(x-\ga)-\gb$ defines an automorphism of $H$ taking $(\ga,\gb)$ to $(0,0)$. Hence at $(\ga,\gb)$, we have
\begin{equation}\label{xbdks}
	y=\gb+\ga^q(x-\ga)+\sum_{i=0}^\infty(-1)^i(x-\ga)^{(q+1)q^i}.
\end{equation}

Now we consider the following problem. Suppose that $Q_i=(\ga_i,\gb_i),1\le i\le r$ are rational points on $H$ with distinct $\ga_i$. Given some positive integers $\mu_i$ for $1\le i\le r$. We want to construct $y-f$ with $f\in\F[x]$ such that $v_{Q_i}(y-f)\ge\mu_i$ for $1\le i\le r$. There are at least two ways to do this.

\paragraph*{First method} For $1\le i\le r$, let $w_i$ be the truncation of the series expansion \eqref{xbdks} of $y$ at $(\ga_i,\gb_i)$ modulo $(x-\ga)^{\mu_i}$, and let $s_i,t_i\in\F[x]$ be defined by
\[
	s_i=\prod_{\substack{j=1\\j\neq i}}^r(x-\ga_j)^{\mu_j}\quad\text{and}\quad s_it_i\equiv 1\mod (x-\ga_i)^{\mu_i}.
\]
Then $y-\sum_{i=1}^r w_is_it_i$ satisfies the required conditions by the Chinese remainder theorem.

\paragraph*{Second method} A somewhat more explicit way is as follows. If $f(x)=\sum_{i=0}^{N-1}a_ix^i\in\F[x]$, then the condition $v_P(y-f)\ge\mu$ is equivalent to the following linear conditions on the coefficients $a_i$,
\[
	\sum_{i=0}^{N-1}\binom{i}{j}\ga^{i-j}a_i=c_j
\]
for $j=0,1,\dots,\mu-1$, where $c_j=0$ except $c_0=\gb$, $c_1=\ga^q$, $c_{(q+1)q^i}=(-1)^i$ for $i\ge 0$. Now let $N=\mu_1+\mu_2+\dots+\mu_r$. Then the required $f$ can be determined by solving the linear system $vA=C$ for the vector $v=(a_0,a_1,a_2,\dots,a_{N-1})$ where $C$ is a certain vector of length $N$ and $A$ is a square matrix of size $N$ obtained by the horizontal join of $N\times\mu_k$ matrices
\[
	A_k=\left[\binom{i}{j}\ga_k^{i-j}\right]_{0\le i\le{N-1},0\le j\le \mu_k-1}
\]
for $1\le k\le r$. The matrix $A$ is called a confluent Vandermonde matrix in the literature, and is known to be invertible (actually the determinant is $\prod_{i,j}(\ga_i-\ga_j)^{\mu_i\mu_j}$ \cite{krattenthaler1999,hou2002}). Therefore the linear system has a unique solution.

\begin{ex}[continued]
Let us compute $f_{1,2}$ in the previous example by the second method. Here $N=1+2+2+3=8$. If $f_{1,2}(x)=\sum_{i=0}^7a_ix^i$, then
$(a_0,a_1,\dots,a_7)A=C$ where
\[
A=\begin{bmatrix}
1 & 1 & 0 & 1 & 0 & 1 & 0 & 0 \\
0 & 1 & 1 & \ga & 1 & \ga^2 & 1 & 0 \\
0 & 1 & 0 & \ga^2 & 0 & \ga & 0 & 1 \\
0 & 1 & 1 & 1 & \ga^2 & 1 & \ga & \ga^2 \\
0 & 1 & 0 & \ga & 0 & \ga^2 & 0 & 0 \\
0 & 1 & 1 & \ga^2 & \ga & \ga & \ga^2 & 0 \\
0 & 1 & 0 & 1 & 0 & 1 & 0 & \ga^2 \\
0 & 1 & 1 & \ga & 1 & \ga^2 & 1 & \ga
\end{bmatrix}
\]
and $C=(1,\ga^2,1,\ga^2,\ga^2,\ga,\ga,0)$. The solution of this linear system was given in the previous subsection.
\end{ex}

\subsection{Converting to a Gr\"obner basis to pick up the $Q$-polynomial}

For this task, we use the Gr\"obner conversion algorithm in \cite{kwankyu2006c} that converts a set of generators of a submodule of $\F[x]^N$ to a module Gr\"obner basis with respect to a special weighted monomial order. We review the algorithm below.

Let $T=\set{(i,j)\mid 0\le i\le l, 0\le j\le q-1}$. Tuples in $T$ are ordered lexicographically such that $(0,0)$ is the first tuple in $T$ and the successor of $(i,j)$ is $(i,j+1)$ if $j<q-1$ or $(i+1,0)$ if $j=q-1$. Thus $\set{y^jz^i\mid (i,j)\in T}$ is a basis for $R[z]_l$ as an $\F[x]$-module and the weight of the basis element $y^jz^i$ is $ui+(q+1)j$. The index of $f\in R[z]_l$ is defined to be the largest tuple $(i,j)$ such that the coefficient of $y^jz^i$ is nonzero. In particular, if the leading term of $f\in R[z]_l$ is $x^iy^jz^k$ with respect to $>_u$, then $\ind(\LT(f))=(k,j)$. Note that $\ind(g_{i,j})=(i,j)$ for the generators $g_{i,j}$ of $I_{M,l}$ computed by Algorithm B.

\begin{algI} 
The algorithm finds the element of $I_{M,l}$ with the smallest leading term. Initially set $g_{i,j}$ to be the initial set of generators of the module $I_{M,l}$ computed by Algorithm B. Let 
\[
	g_{i,j}=\sum_{(i',j')\in T}a_{i,j,i',j'}y^{j'}z^{i'}
\]
during the execution of the algorithm. For $r=(r_1,r_2)$ and $s=(s_1,s_2)$ in $T$, we abbreviate $a_{r,s}=a_{r_1,r_2,s_1,s_2}$. 
\begin{itemize}
\item[I1.] Set $r\leftarrow (0,0)$.
\item[I2.] Set $r$ to the successor of $r$. If $r\in T$, then proceed; otherwise go to step I6.
\item[I3.] Set $s\leftarrow \ind(\LT(g_r))$. If $s=r$, then go to step I2.
\item[I4.] Set $d\leftarrow\deg(a_{r,s})-\deg(a_{s,s})$ and  $c\leftarrow\LC(a_{r,s})\LC(a_{s,s})^{-1}$.
\item[I5.] If $d\ge 0$, then set 
\[
	g_r\leftarrow g_r-cx^dg_s.
\]
If $d<0$, then set, storing $g_s$ in a temporary variable,
\[
	g_s\leftarrow g_r,\quad g_r\leftarrow x^{-d}g_r-cg_s.
\]
Go back to step I3.
\item[I6.] Output $g_{i,j}$ with the smallest leading term, and the algorithm terminates.  
\end{itemize}
\end{algI}

\begin{ex}[continued]
Algorithm I converts the initial basis given in \eqref{dwhxas} to a Gr\"obner basis with respect to the order $>_u$. The computed Gr\"obner basis is
\[
\begin{bmatrix}
\ga x^{10}+\cdots & \ga^2x^8+\cdots & \ga x^7+\cdots & x^5+\cdots & \cdots & 0 & 0 & 0 & 0 \\
x^{10}+\cdots & \ga x^9+\cdots & \ga x^8+\cdots & \ga x^6+\cdots & \cdots & 0 & 0 & 0 & 0 \\
\ga x^{10}+\cdots & \ga x^8+\cdots & \ga^2x^8+\cdots & x^6+\cdots & \cdots & 0 & 0 & 0 & 0 \\
x^8+\cdots & \ga^2x^8+\cdots & x^7+\cdots & x^6+\cdots & \cdots & 0 & 0 & 0 & 0 \\
x^9+\cdots & x^7+\cdots & x^6+\cdots & x^5+\cdots & \cdots & 0 & 0 & 0 & 0 \\
x^9+\cdots & \ga^2x^8+\cdots & \ga x^7+\cdots & \ga x^6+\cdots & \cdots & 0 & 0 & 0 & 0 \\
\ga x^8+\cdots & \ga x^6+\ga x^3 & \ga^2x^7+\cdots & \ga^2x^5+\cdots & \cdots & 0 & 0 & 0 & 0 \\
x^8+\cdots & x^6+\cdots & \ga^2x^5+\cdots & \ga^2x^5+\cdots & \cdots & 1 & 0 & 0 & 0 \\
\ga^2x^9+\cdots & \ga^2x^7+\cdots & x^7+\cdots & \ga x^5+\cdots & \cdots & x & 0 & 0 & 0 \\
x^9+\cdots & \ga x^7+\cdots & \ga x^7+\cdots & x^6+\cdots & \cdots & 0 & 1 & 0 & 0 \\
\ga x^{10}+\cdots & \ga x^8+\cdots & x^8+\cdots & \ga x^6+\cdots & \cdots & \ga^2x^2+\ga^2x & \ga^2 & 1 & 0 \\
\ga^2x^{11}+\cdots & x^{10}+\cdots & \ga x^9+\cdots & x^8+\cdots & \cdots & \ga^2x^3 & x^2+\cdots & 0 & 1
\end{bmatrix}.
\]
The twelve rows represent the polynomials in the Gr\"obner basis of the module $I_{M,l}$ over $\F[x]$. Comparing the weights of the leading coefficients of the polynomials, which lie on the diagonal, we find that the polynomial represented by the eleventh row is the required $Q$-polynomial of the ideal $I_M$, given explicitly in Example \ref{cjops} equation \eqref{dwkxzs}.
\end{ex}

\section{Complexity Analysis}

Elements of $R$ can be written uniquely as polynomials in $y$ of degree less than $q$ with coefficients in $\F[x]$. We assume that for computations in $R$, we use this representation of elements of $R$. Also we think of $\deg_x(f)$ and $\deg_y(f)$ for $f\in R$ in this representation. Note that a straightforward way of multiplying two elements $f,g$ of $R$ takes $O(q^2ab)$ multiplications on $\F$ and that $\deg_x(fg)\le a+b+q+1$ if $a=\deg_x(f)$ and $b=\deg_x(g)$.

First we consider computing $f\in\F[x]$ satisfying $v_{P_i}(y-f)\ge\mu_i$ for $1\le i\le r$ as in Section \ref{subsecC}. This computation takes $O(N^3)$ multiplications on $\F$ where $N=\mu_1+\mu_2+\dots+\mu_r$, if we use Gaussian elimination to solve the linear system. Note also $\deg_x(f)\le N-1$.

Next we consider computing $g_c$ according to Proposition \ref{prop6} in Section \ref{subsecB}. The first product $\pi_1$ on the right side of \eqref{equxcjms} has at most $lq^2$ linear factors. Hence $\pi_1$ can be computed with $O(l^2q^4)$ multiplications on $\F$. Note $\deg_x(\pi_1)\le lq^2$. On the other hand, as $\deg_x(f_{b,c})<lq^2$, the second product $\pi_2$ can be computed with $O(c^2l^2q^4)$ multiplications on $\F$. Note $\deg_y(\pi_2)\le c-1$ and $\deg_x(\pi_2)\le(c-1)lq^2$. Then $\pi_1$ and $\pi_2$ can be multiplied with $O(c^2l^2q^4)$ multiplications on $\F$. Hence, in total, computing $g_c$ takes $O(c^2l^2q^4)$ multiplications on $\F$. Note $\deg_x(g_c)\le clq^2$ and $\deg_y(g_c)\le c-1$.

Now we consider computations in steps B2 and B3 of Algorithm B in Section \ref{subsecA}. Fix $s$. Computing $\eta_t$ ($=g_{t+1}$), as shown above, takes $O((t+1)^2l^2q^4)$ multiplications on $\F$ for each $t=0,1,\dots,q-1$. Computing $h^{(s)}$ can be done with $O(nq^2)$ multiplications on $\F$. Note $\deg_x(h^{(s)})\le q^2-1$. Let $\pi^{(s)}$ denote the product of the right side in \eqref{eqjcjsq}. It is easy to verify $\deg_z(\pi^{(s)})=s$ and $\deg_x(\pi^{(s)})\le s(q^2-1)+(s-1)(q+1)$ if $s\ge 1$. So computing $g_{s,t}=\eta_t\pi^{(s)}$ takes $O(s^2tlq^6)$ multiplications on $\F$. Note $\deg_x(g_{s,t})\le tlq^2+s(q^2+q)$. Computing $\pi^{(s+1)}=\pi^{(s)}(z-h^{(s)})$ takes $O(s^2q^6)$ multiplications on $\F$.

Summing up, an execution of Algorithm B takes
\[
\begin{split}
	&\sum_{s=0}^l\left(O(nq^2)+O(s^2q^6)+\sum_{t=0}^{q-1}O((t+1)^2l^2q^4)+\sum_{t=0}^{q-1}O(s^2tlq^6)\right)\\
	&\quad=\sum_{s=0}^l\left(O(s^2q^6)+O(l^2q^7)+O(s^2lq^8)\right)\\
	&\quad=O(l^3q^6)+O(l^3q^7)+O(l^4q^8)\\
	&\quad=O(l^4q^8)
\end{split}
\]
multiplications on $\F$. Lastly noting $\deg_u(g_{s,t})=O(lq^4)$ and using a result in \cite{kwankyu2006c}, we see that an execution of Algorithm I takes $O(l^5q^{10})$ multiplications on $\F$. Therefore the algebraic soft-decision decoder of Hermitian codes can be implemented in a way that takes $O(l^5q^{10})=O(l^5n^{3+1/3})$ multiplications on $\F$.

\section{Simulation results}

We implemented the algebraic soft-decision decoder (SDD) for Hermitian codes in software. In this section, we present some simulation results that show the performance of the algebraic soft-decision decode for half-rate Hermitian codes. 

First we describe the general setup of our simulations. We assume the AWGN channel. For QPSK and QAM modulations, the signal points correspond one-to-one with the symbols in the finite field over which the code is defined, and the posterior probabilities of the symbols are simply set to those of the corresponding signal points. For BPSK, each of the symbols correspond with a bit sequence, and the posterior probabilities of the symbols are set to the products of the posterior probabilities of the bits. Koetter and Vardy's multiplicity assignment algorithm \cite{koetter2003} is used to translate the posterior probabilities to the values of the multiplicity matrix. The multiplicity assignment algorithm accepts a parameter $L$ that limits the $z$-degree of the $Q$-polynomial, thereby the list size of the candidate codewords to at most $L$. From the multiplicity matrix, our interpolation algorithm finds the $Q$-polynomial. Then Wu and Siegel's root-finding algorithm \cite{wu2001} is used to compute the roots of the $Q$-polynomial. The list of candidate codewords is then formed from the roots. If the list is empty, then the decoder simply output the message part of the received vector determined by hard-decision directly from the posterior probabilities of the symbols. If the list is not empty, the decoder outputs the message from the codeword that has the best score with respect to the multiplicity matrix. 

\subsection{$[8,4]$ Hermitian code with QPSK}

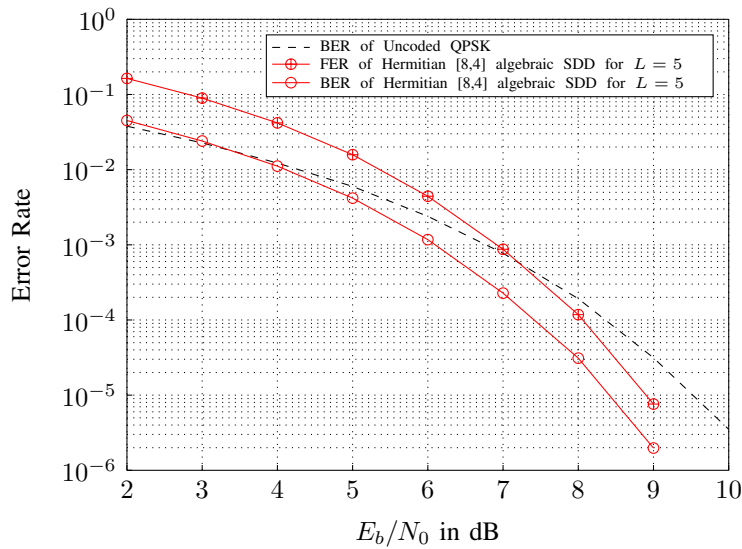
\begin{figure}
\begin{center}
\beginpgfgraphicnamed{pic1}
\begin{tikzpicture}
\begin{scope}[xscale=1]
	\draw[-] (2,0) -- (10,0);
	\draw[-] (2,6) -- (10,6);
	\draw[-] (2,0) -- (2,6);
	\draw[-] (10,0) -- (10,6);
	\foreach \y in {-1, ..., -6}
		\foreach \inc in {0.301, 0.477, 0.602, 0.698, 0.778, 0.845, 0.903, 0.954,1}
			\draw[dotted,yshift=6cm][yshift=\inc cm] (2,\y) -- (10,\y);
	\foreach \x in {2, ..., 10}
		\draw[dotted] (\x,0) -- (\x,6);
	\foreach \x/\xtext in {2, ..., 10}
		\draw[shift={(\x,0)}] (0pt,2pt) -- (0pt,0pt) node[below] {\small $\xtext$};
	\foreach \y/\ytext in {0, ..., -6}
		\draw[shift={(2,\y)}] [yshift=6cm] (2pt,0pt) -- (0pt,0pt) node[left] {\small $10^{\ytext}$};
	\draw[dashed] plot[yshift=6cm] file{awgn-qpsk-hd-uncoded-BER.table};
	\draw[red] plot[mark=oplus,yshift=6cm] file{awgn-qpsk-sd-Hermitian[8,4]-ASD-L5-FER.table};
	\draw[red] plot[mark=o,yshift=6cm] file{awgn-qpsk-sd-Hermitian[8,4]-ASD-L5-BER.table};
	\end{scope}
	\draw (9.8,5.8) node[legend,text width=15em] {
		\tiny \tikz[baseline=-.5ex] \draw[dashed] plot coordinates {(-.5,0)(0,0)}; 
			BER of Uncoded QPSK \\
		\tiny \tikz \draw[red] plot[mark=oplus,mark indices={2}] coordinates {(-.5,0)(-.25,0)(0,0)}; 
			FER of Hermitian [8,4] algebraic SDD for $L=5$\\
		\tiny \tikz \draw[red] plot[mark=o,mark indices={2}] coordinates {(-.5,0)(-.25,0)(0,0)}; 
			BER of Hermitian [8,4] algebraic SDD for $L=5$\\
	};
	\draw (6,0) [yshift=-15pt] node[below] {\small $E_b/N_0$ in dB};
	\draw (2,3) [xshift=-40pt] node[rotate=90] {\small Error Rate};
\end{tikzpicture}
\endpgfgraphicnamed
\end{center}
\caption{\small\label{fig1} Performance of algebraic SDD of $[8,4]$ Hermitian code over $\F_4$ with QPSK modulation.}
\end{figure}

The smallest field over which Hermitian codes are defined is $\F_4$ and the length of these codes is $8$. The length is extremely small, and it is perhaps unrealistic to expect the codes to be used in practice. However the codes are amenable for simulations with somewhat larger $L$. Figure \ref{fig1} show the performance of the half-rate $[8,4]$ Hermitian code with QPSK. The example used in previous sections was sampled from this simulation with SNR $2$ and $L=5$.

\subsection{$[64,32]$ Hermitian code with BPSK}

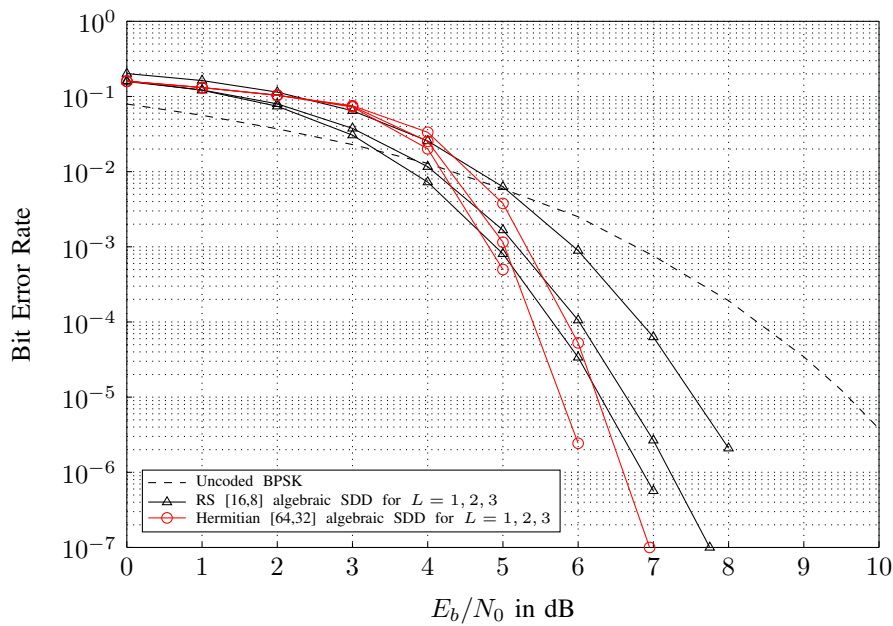
\begin{figure}
\begin{center}
\beginpgfgraphicnamed{pic2}
\begin{tikzpicture}
\begin{scope}[xscale=1,yshift=7cm]
	\draw[-] (0,0) -- (10,0);
	\draw[-] (0,-7) -- (10,-7);
	\draw[-] (0,0) -- (0,-7);
	\draw[-] (10,0) -- (10,-7);
	\foreach \y in {-1, ..., -7}
		\foreach \inc in {0.301, 0.477, 0.602, 0.698, 0.778, 0.845, 0.903, 0.954,1}
			\draw[dotted][yshift=\inc cm] (0,\y) -- (10,\y);
	\foreach \x in {0, ..., 10}
		\draw[dotted] (\x,0) -- (\x,-7);
	\foreach \x/\xtext in {0, 1, ..., 10}
		\draw[shift={(\x,-7)}] (0pt,2pt) -- (0pt,0pt) node[below] {\small $\xtext$};
	\foreach \y/\ytext in {0, ..., -7}
		\draw[shift={(0,\y)}] (2pt,0pt) -- (0pt,0pt) node[left] {\small $10^{\ytext}$};
	\draw[dashed] plot file{awgn-bpsk-hd-uncoded.table};
	\draw plot[mark=triangle] file{awgn-bpsk-sd-RS[16,8]-ASD-L1-BER.table};
	\draw plot[mark=triangle] file{awgn-bpsk-sd-RS[16,8]-ASD-L2-BER.table};
	\draw plot[mark=triangle] file{awgn-bpsk-sd-RS[16,8]-ASD-L3-BER.table};
	\draw[red] plot[mark=o] file{awgn-bpsk-sd-Hermitian[64,32]-ASD-L1-BER.table};
	\draw[red] plot[mark=o] file{awgn-bpsk-sd-Hermitian[64,32]-ASD-L2-BER.table};
	\draw[red] plot[mark=o] file{awgn-bpsk-sd-Hermitian[64,32]-ASD-L3-BER.table};
\end{scope}
	\draw (.2,.2) node[legendsw,text width=14em] {
		\tiny \tikz[baseline=-.5ex] \draw[dashed] plot coordinates {(-.5,0)(0,0)}; 
			Uncoded BPSK \\
		\tiny \tikz \draw plot[mark=triangle,mark indices={2}] coordinates {(-.5,0)(-.25,0)(0,0)}; 
			RS [16,8] algebraic SDD for $L=1,2,3$\\
		\tiny \tikz \draw[red] plot[mark=o,mark indices={2}] coordinates {(-.5,0)(-.25,0)(0,0)}; 
			Hermitian [64,32] algebraic SDD for $L=1,2,3$\\
	};
	\draw (5,0) [yshift=-15pt] node[below] {\small $E_b/N_0$ in dB};
	\draw (0,3.5) [xshift=-40pt] node[rotate=90] {\small Bit Error Rate};
\end{tikzpicture}
\endpgfgraphicnamed
\end{center}
\caption{\small\label{fig3} Bit Error Performance of algebraic SDD of $[64,32]$ Hermitian code over $\F_{16}$ with BPSK modulation.}
\end{figure}

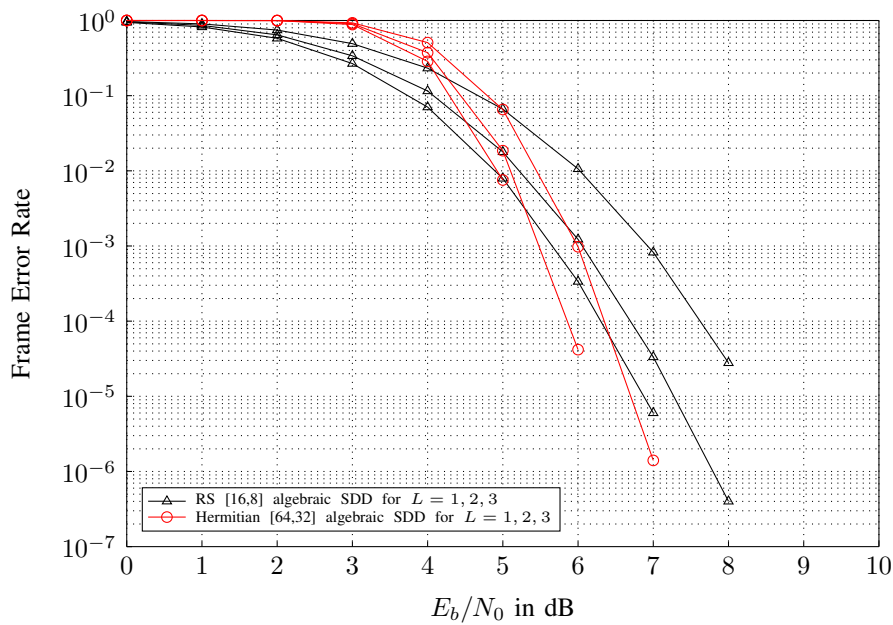
\begin{figure}
\begin{center}
\beginpgfgraphicnamed{pic3}
\begin{tikzpicture}
\begin{scope}[xscale=1,yshift=7cm]
	\draw[-] (0,0) -- (10,0);
	\draw[-] (0,-7) -- (10,-7);
	\draw[-] (0,0) -- (0,-7);
	\draw[-] (10,0) -- (10,-7);
	\foreach \y in {-1, ..., -7}
		\foreach \inc in {0.301, 0.477, 0.602, 0.698, 0.778, 0.845, 0.903, 0.954,1}
			\draw[dotted][yshift=\inc cm] (0,\y) -- (10,\y);
	\foreach \x in {0, ..., 10}
		\draw[dotted] (\x,0) -- (\x,-7);
	\foreach \x/\xtext in {0, 1, ..., 10}
		\draw[shift={(\x,-7)}] (0pt,2pt) -- (0pt,0pt) node[below] {\small $\xtext$};
	\foreach \y/\ytext in {0, ..., -7}
		\draw[shift={(0,\y)}] (2pt,0pt) -- (0pt,0pt) node[left] {\small $10^{\ytext}$};
	\draw plot[mark=triangle] file{awgn-bpsk-sd-RS[16,8]-ASD-L1-FER.table};
	\draw plot[mark=triangle] file{awgn-bpsk-sd-RS[16,8]-ASD-L2-FER.table};
	\draw plot[mark=triangle] file{awgn-bpsk-sd-RS[16,8]-ASD-L3-FER.table};
	\draw[red] plot[mark=o] file{awgn-bpsk-sd-Hermitian[64,32]-ASD-L1-FER.table};
	\draw[red] plot[mark=o] file{awgn-bpsk-sd-Hermitian[64,32]-ASD-L2-FER.table};
	\draw[red] plot[mark=o] file{awgn-bpsk-sd-Hermitian[64,32]-ASD-L3-FER.table};
\end{scope}
	\draw (.2,.2) node[legendsw,text width=14em] {
		\tiny \tikz \draw plot[mark=triangle,mark indices={2}] coordinates {(-.5,0)(-.25,0)(0,0)}; 
			RS [16,8] algebraic SDD for $L=1,2,3$\\
		\tiny \tikz \draw[red] plot[mark=o,mark indices={2}] coordinates {(-.5,0)(-.25,0)(0,0)}; 
			Hermitian [64,32] algebraic SDD for $L=1,2,3$\\
	};
	\draw (5,0) [yshift=-15pt] node[below] {\small $E_b/N_0$ in dB};
	\draw (0,3.5) [xshift=-40pt] node[rotate=90] {\small Frame Error Rate};
\end{tikzpicture}
\endpgfgraphicnamed
\end{center}
\caption{\small\label{fig4} Frame Error Performance of algebraic SDD of $[64,32]$ Hermitian code over $\F_{16}$  with BPSK modulation.}
\end{figure}

Figures \ref{fig3} and \ref{fig4} show the performance of $[64,32]$ Hermitian code with BPSK modulation. For comparison, the figures also show the performance of the half-rate $[16,8]$ Reed-Solomon code. Observe that the performance curve of Hermitian code more steeply decrease than that of Reed-Solomon code, and from around 5 dB, the Hermitian code outperforms the Reed-Solomon code.

\subsection{$[64,32]$ Hermitian code with $16$-QAM}

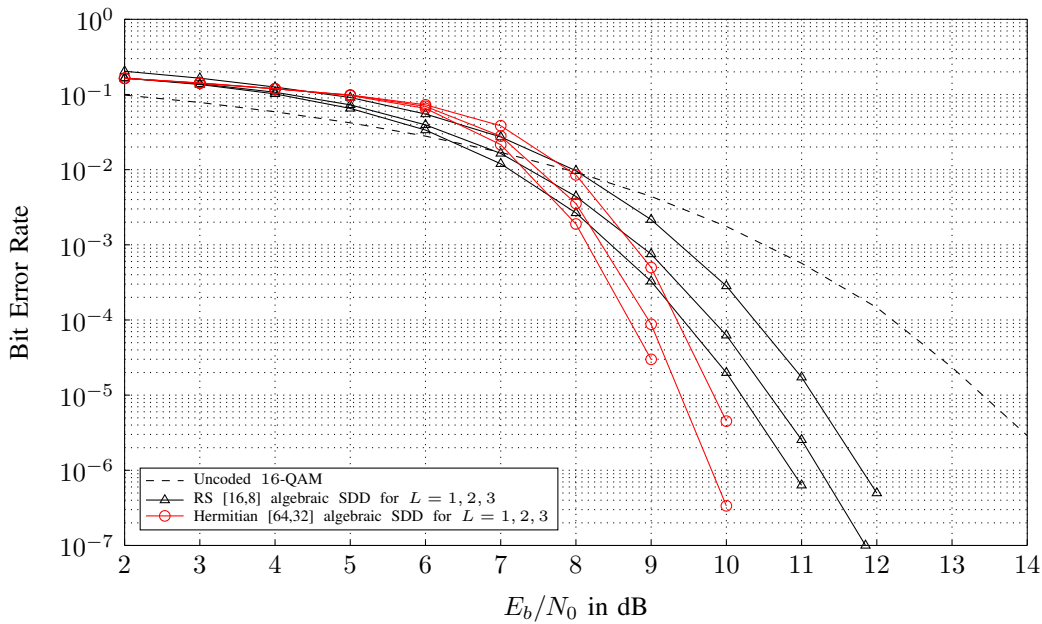
\begin{figure}
\begin{center}
\beginpgfgraphicnamed{pic4}
\begin{tikzpicture}
\begin{scope}[xscale=1,yshift=7cm]
	\draw[-] (2,0) -- (14,0);
	\draw[-] (2,-7) -- (14,-7);
	\draw[-] (2,0) -- (2,-7);
	\draw[-] (14,0) -- (14,-7);
	\foreach \y in {-1, ..., -7}
		\foreach \inc in {0.301, 0.477, 0.602, 0.698, 0.778, 0.845, 0.903, 0.954,1}
			\draw[dotted][yshift=\inc cm] (2,\y) -- (14,\y);
	\foreach \x in {2, ..., 14}
		\draw[dotted] (\x,0) -- (\x,-7);
	\foreach \x/\xtext in {2, ..., 14}
		\draw[shift={(\x,-7)}] (0pt,2pt) -- (0pt,0pt) node[below] {\small $\xtext$};
	\foreach \y/\ytext in {0, ..., -7}
		\draw[shift={(2,\y)}] (2pt,0pt) -- (0pt,0pt) node[left] {\small $10^{\ytext}$};
	\draw[dashed] plot file{awgn-16qam-hd-uncoded-BER.table};
	\draw plot[mark=triangle] file{awgn-16qam-sd-RS[16,8]-ASD-L1-BER.table};
	\draw plot[mark=triangle] file{awgn-16qam-sd-RS[16,8]-ASD-L2-BER.table};
	\draw plot[mark=triangle] file{awgn-16qam-sd-RS[16,8]-ASD-L3-BER.table};
	\draw[red] plot[mark=o] file{awgn-16qam-sd-Hermitian[64,32]-ASD-L1-BER.table};
	\draw[red] plot[mark=o] file{awgn-16qam-sd-Hermitian[64,32]-ASD-L2-BER.table};
	\draw[red] plot[mark=o] file{awgn-16qam-sd-Hermitian[64,32]-ASD-L3-BER.table};
\end{scope}
	\draw (2.2,.2) node[legendsw,text width=14em] {
		\tiny \tikz[baseline=-.5ex] \draw[dashed] plot coordinates {(-.5,0)(0,0)}; 
			Uncoded $16$-QAM \\
		\tiny \tikz \draw plot[mark=triangle,mark indices={2}] coordinates {(-.5,0)(-.25,0)(0,0)}; 
			RS [16,8] algebraic SDD for $L=1,2,3$\\
		\tiny \tikz \draw[red] plot[mark=o,mark indices={2}] coordinates {(-.5,0)(-.25,0)(0,0)}; 
			Hermitian [64,32] algebraic SDD for $L=1,2,3$\\
	};
	\draw (8,0) [yshift=-15pt] node[below] {\small $E_b/N_0$ in dB};
	\draw (2,3.5) [xshift=-40pt] node[rotate=90] {\small Bit Error Rate};
\end{tikzpicture}
\endpgfgraphicnamed
\end{center}
\caption{\small\label{fig5} Bit Error Performance of algebraic SDD of $[64,32]$ Hermitian code with $16$-QAM modulation.}
\end{figure}

\begin{figure}
\begin{center}
\beginpgfgraphicnamed{pic5}
\begin{tikzpicture}
\begin{scope}[xscale=1,yshift=7cm]
	\draw[-] (2,0) -- (14,0);
	\draw[-] (2,-7) -- (14,-7);
	\draw[-] (2,0) -- (2,-7);
	\draw[-] (14,0) -- (14,-7);
	\foreach \y in {-1, ..., -7}
		\foreach \inc in {0.301, 0.477, 0.602, 0.698, 0.778, 0.845, 0.903, 0.954,1}
			\draw[dotted][yshift=\inc cm] (2,\y) -- (14,\y);
	\foreach \x in {2, ..., 14}
		\draw[dotted] (\x,0) -- (\x,-7);
	\foreach \x/\xtext in {2, ..., 14}
		\draw[shift={(\x,-7)}] (0pt,2pt) -- (0pt,0pt) node[below] {\small $\xtext$};
	\foreach \y/\ytext in {0, ..., -7}
		\draw[shift={(2,\y)}] (2pt,0pt) -- (0pt,0pt) node[left] {\small $10^{\ytext}$};
	\draw plot[mark=triangle] file{awgn-16qam-sd-RS[16,8]-ASD-L1-FER.table};
	\draw plot[mark=triangle] file{awgn-16qam-sd-RS[16,8]-ASD-L2-FER.table};
	\draw plot[mark=triangle] file{awgn-16qam-sd-RS[16,8]-ASD-L3-FER.table};
	\draw[red] plot[mark=o] file{awgn-16qam-sd-Hermitian[64,32]-ASD-L1-FER.table};
	\draw[red] plot[mark=o] file{awgn-16qam-sd-Hermitian[64,32]-ASD-L2-FER.table};
	\draw[red] plot[mark=o] file{awgn-16qam-sd-Hermitian[64,32]-ASD-L3-FER.table};
\end{scope}
	\draw (2.2,.2) node[legendsw,text width=14em] {
		\tiny \tikz \draw plot[mark=triangle,mark indices={2}] coordinates {(-.5,0)(-.25,0)(0,0)}; 
			RS [16,8] algebraic SDD for $L=1,2,3$\\
		\tiny \tikz \draw[red] plot[mark=o,mark indices={2}] coordinates {(-.5,0)(-.25,0)(0,0)}; 
			Hermitian [64,32] algebraic SDD for $L=1,2,3$\\
	};
	\draw (8,0) [yshift=-15pt] node[below] {\small $E_b/N_0$ in dB};
	\draw (2,3.5) [xshift=-40pt] node[rotate=90] {\small Frame Error Rate};
\end{tikzpicture}
\endpgfgraphicnamed
\end{center}
\caption{\small\label{fig6} Frame Error Performance of algebraic SDD of $[64,32]$ Hermitian code with $16$-QAM modulation.}
\end{figure}
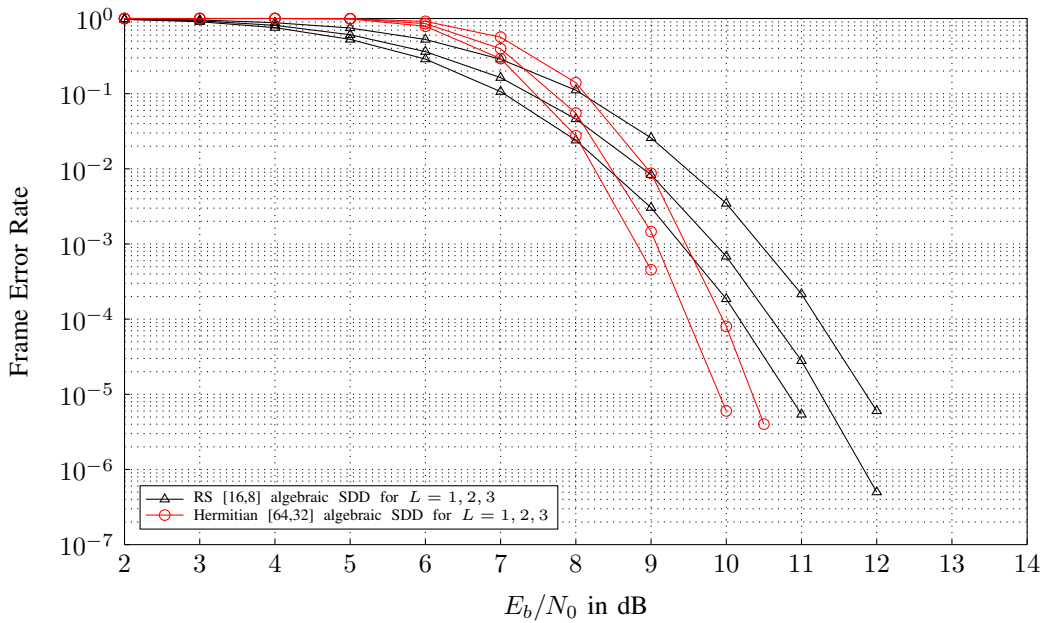

Figures \ref{fig5} and \ref{fig6} also show that the Hermitian code outperforms the Reed-Solomon code with $16$-QAM modulation, from around 8 dB onward.

\section{Conclusion}

We presented an algebraic soft-decision decoder for Hermitian codes. Software simulations show that Hermitian codes perform better than Reed-Solomon codes for algebraic soft-decision decoding, as expected. However, for the decoder to be really practical, reduction of the computational complexity remains an important problem. One promising avenue is to generalize the idea of complexity reduction for Reed-Solomon codes in \cite{ma2007a}. Designing efficient electric circuits implementing the decoder is of course an issue to explore.

The extent of our computer simulations of the decoding algorithm was limited by our computing resources. It would be good to have analytic results about the performance of the decoding algorithm. There have been several analytic performance analyses for the algebraic soft-decision decoding of Reed-Solomon codes \cite{ratnakar2005}. Similar analyses may be done for Hermitian codes. 

Our description of the decoding algorithm is interwoven with the particular structure of Hermitian codes. However, the underlying principle of the decoding algorithm seems to apply to a wider class of algebraic geometry codes. In particular, plane algebraic curves with one point at infinity are immediate candidates. We leave an adequate treatment of this subject as a remaining work.

%


\end{document}